\documentclass[sn-mathphys,Numbered]{sn-jnl}

\usepackage{graphicx}%
\usepackage{multirow}%
\usepackage{amsmath,amssymb,amsfonts}%
\usepackage{amsthm}%
\usepackage{mathrsfs}%
\usepackage[title]{appendix}%
\usepackage{xcolor}%
\usepackage{textcomp}%
\usepackage{manyfoot}%
\usepackage{booktabs}%
\usepackage{algorithm}%
\usepackage{algorithmicx}%
\usepackage{algpseudocode}%
\usepackage{listings}%

\theoremstyle{thmstyleone}%
\theoremstyle{thmstyletwo}%

\theoremstyle{thmstylethree}%

\begin{document}

\title[Article Title]{Towards fully integrated photonic backpropagation training and inference using on-chip nonlinear activation and gradient functions}

\author*[1]{\fnm{Farshid} \sur{Ashtiani}}\email{farshid.ashtiani@nokia-bell-labs.com}

\author[1]{\fnm{Mohamad Hossein} \sur{Idjadi}}

\affil*[1]{\orgname{Nokia Bell Labs}, \orgaddress{\street{600 Mountain Ave}, \city{Murray Hill}, \postcode{07974}, \state{NJ}, \country{USA}}}


\abstract{Gradient descent-based backpropagation training is widely used in many neural network systems. However, photonic implementation of such method is not straightforward mainly since having both the nonlinear activation function and its gradient using standard integrated photonic components is challenging. Here, we demonstrate the realization of two commonly used neural nonlinear activation functions and their gradients on a silicon photonic platform. Our method leverages the nonlinear electro-optic response of a micro-disk modulator. As a proof of concept, the experimental results are incorporated into a neural network simulation platform to classify MNIST handwritten digits dataset where we classification accuracies of more than 97\% are achieved that are on par with those of ideal nonlinearities and gradients.}




\maketitle

\section{Introduction}\label{sec1}

As artificial neural networks (ANN) are being utilized in a wider variety of applications from natural language models to medical diagnosis \cite{nlp,med}, the need for higher processing speed and lower energy requirements is increasing. In supervised learning, ANNs are first trained on a specific dataset to "learn" a certain task (training), and then perform the same task on an unseen dataset (inference). To learn the optimized network parameters (\textit{i.e.}, weights and biases), a training algorithm such as gradient descent-based backpropagation (BP) \cite{dl} is used. It is an iterative process where both linear (weight and sum) and nonlinear (activation) computations are performed many times to reach an optimized solution. Therefore, improving speed and energy efficiency of such a computation intensive process can significantly benefit the overall performance of a neural network.

Benefiting from low-loss interconnects and a wide available bandwidth at optical frequencies, photonic neural networks (PNN) have become a promising candidate to augment the performance of conventional digital clock-based processors. Such photonic analog processors can process optical signals as they propagate through the system which enables processing at the speed of light. Several PNNs have been demonstrated for different applications such as optical fiber nonlinearity compensation \cite{fbr_nonlin} and image classification \cite{image,image2}. Despite promising results, many of these systems rely on digital computers for offline BP training and show inference using photonic chips. One main reason is that for the BP algorithm to work, the nonlinear activation function (in the forward path) and its gradient (in the backward bath) are required to calculate the proper adjustments of the weights and biases that minimize the cost/error function. Since photonic implementation of such asymmetric nonlinearity is not straightforward \cite{prospect}, alternative photonic-friendly training algorithms such as finite difference method \cite{fdm}, direct feedback alignment \cite{dfa}, and simultaneous perturbation method \cite{perturb} have been proposed and demonstrated. Nevertheless, due to the wide applicability and a large number of existing neural network models based on BP training, together with the great potential of photonic computing, photonic implementation of BP training can be very beneficial. Recently, a hybrid digital-photonic demonstration of BP has been reported \cite{bp_mz}. Despite impressive results, only linear computation is realized photonically, and the nonlinear activation and its gradient are implemented digitally. 

\begin{figure}[h!]%
\centering
\includegraphics[width=1\textwidth]{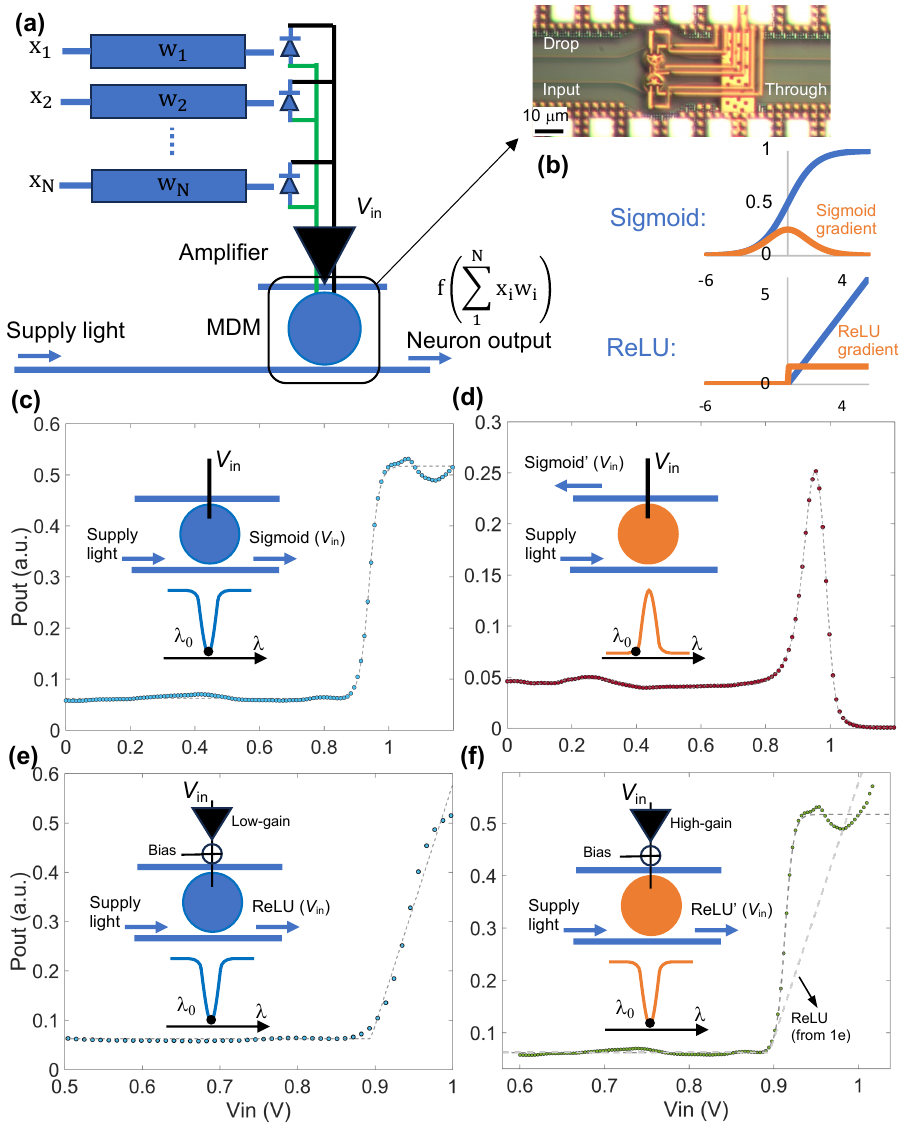}
\caption{(a) Photonic neuron with OEO activation (b) Ideal sigmoid, ReLU, and their corresponding gradients. Proposed implementations of (c) sigmoid, (d) sigmoid gradient, (e) ReLU, and (f) ReLU gradient. The corresponding measured data (points) and the fitted curves (dashed lines) are shown for each case.}\label{fig1}
\end{figure}

Here, we propose a novel photonic-friendly solution to realize two commonly used nonlinear activation functions, namely sigmoid and rectified linear unit (ReLU), and their corresponding gradients on a standard silicon photonic platform to enable fully integrated photonic BP training and inference. We leverage the nonlinear electro-optic response of an add-drop micro-disk modulator (MDM) to closely approximate those functions. We then use the measured data in a neural network simulation platform with customized activation and gradient functions to demonstrate the applicability of the proposed method to classify MNIST handwritten digits dataset. Accuracies of 97\% and 98\% are achieved using the approximated sigmoid and ReLU function and gradient, respectively, that are on par with those of ideal activation functions. The chip was fabricated in AIM Photonics 180 nm silicon photonic process and a microphotograph is shown in Fig. \ref{fig1}a.

\section{Activation function and gradient approximation using a micro-disk modulator}\label{sec2}

Optical-electrical-optical (OEO) conversion enables high-speed and scalable neural nonlinearity \cite{image}. Figure \ref{fig1}a shows a typical application of OEO nonlinearity, f(.), in a photonic neuron. The relative weights between the optical inputs ($x_{i}$) are set using intensity modulators ($w_{i}$), and the weighted-sum of the inputs is generated using parallel photodiodes. The photocurrent drives the PN-junction of a MDM to realize the nonlinear activation function by modulating an independent supply light to generate the optical output of the neuron.

\subsection{Sigmoid function }

Sigmoid is a commonly used function due to having a smooth gradient and a normalized output (\textit{i.e.}, between 0 and 1). Figure \ref{fig1}b shows sigmoid and its gradient. We propose the schemes in Figs. \ref{fig1}c and \ref{fig1}d to approximate sigmoid and its gradient. For the sigmoid function, the MDM resonance is thermally tuned to align to the supply light wavelength ($\lambda_{0}$=1539.15 nm) and the through port is monitored. Once $V_{in}$ exceeds the turn-on voltage of the PN-junction (about 0.9V), the MDM resonance shifts and the output power increases. The output power flattens once the supply light wavelength is well beyond the shifted MDM resonance, resulting in a sigmoid-like function. The measured data and the corresponding sigmoid fit are shown in Fig. \ref{fig1}c. To approximate the gradient, the drop port is used. The MDM resonance (in BP mode) is biased at $\lambda_{0}$, and then thermally tuned until the drop port output falls below a certain threshold (Fig. \ref{fig1}d). Once biased, as $V_{in}$ increases a sigmoid gradient-like response is achieved (Fig. \ref{fig1}d). Note that the output offset for $V_{IN} < 0.9 V$ is function of the threshold voltage in the alignment process and can be adjusted.

\subsection{ReLU function }

ReLU is another commonly used neural nonlinearity that due to the simplicity of the function and its gradient offers computational efficiency and also features non-vanishing gradient, resulting in fast convergence (Fig. \ref{fig1}b). Figure \ref{fig1}e and \ref{fig1}f show the architecture to approximate ReLU and its gradient. ReLU approximation is similar to case of sigmoid, except in this case the maximum of $V_{in}$ is intentionally limited by controlling the electronic gain and a bias voltage to avoid flattening at higher values of $V_{in}$ (low-gain mode). To approximate its gradient which is a step function, electronic gain is increased (high-gain mode) which makes the transition to the flat region very sharp (\textit{i.e.}, approximating a step function). Note that the threshold voltage of $0.9 V$ is kept the same via the bias voltage. The measured ReLU, its gradient, and the corresponding fitted curves are shown in Fig. 1e and 1f. Next, the measurement results are incorporated into a neural network simulation to validate the accuracy of the proposed method. 

\section{MNIST classification results}\label{sec3}

\begin{figure}[h]%
\centering
\includegraphics[width=0.95\textwidth]{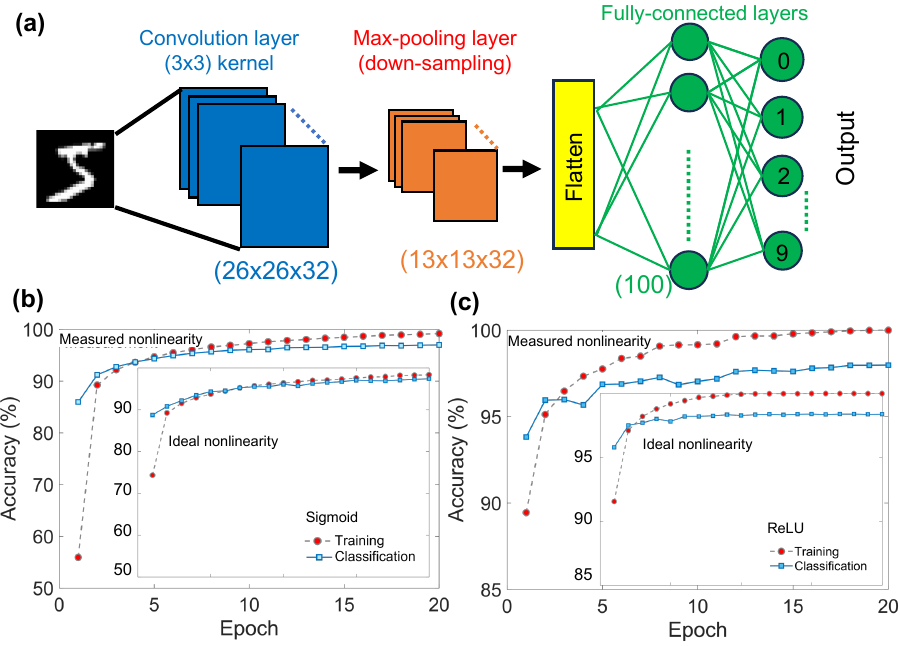}
\caption{(a) Neural network model for MNIST classification. The convolution and the first fully-connected neurons use the measured data as their activation function (for inference) and gradient (for training). This is achieved by defining custom activation and gradient functions in the neural network model. Training and classification accuracy comparison between the scenarios that ideal and measured activation functions and gradients for (b) sigmoid, and (c) ReLU are used.}\label{fig2}
\end{figure}

The measured results (fitted curves) for both sigmoid and ReLU functions and gradients are used in a neural network model to classify MNIST handwritten digits dataset. The architecture of the network is shown in Fig. \ref{fig2}a. The network consists of a 2-D convolution layer, a 2D max-pooling layer, and two fully-connected layers with 100 and 10 neurons, respectively, to generate the classification result. Note that custom activation and gradient functions are incorporated into the model in order to use the actual measured data in the convolution and first fully-connected neurons. The output layer uses softmax activation. 

For training, stochastic gradient descent BP with a learning rate of 0.01 and momentum of 0.9 is used. He-uniform and categorical cross-entropy are used for kernel initialization and loss function, respectively. Figures \ref{fig2}b and \ref{fig2}c compare the training and classification accuracy for ideal and measured sigmoid and ReLU functions and gradients, respectively. Classification accuracies of 97\% and 98\% are achieved when measured data for sigmoid and ReLU are used in the model, respectively. We achieve very similar results when the measured data replaces the ideal functions which shows that our proposed method, especially for gradient calculation, is accurate enough and does not affect the performance of the neural network.

\section{Conclusion}\label{sec4}

We demonstrated the first implementation of the gradient of OEO neural nonlinearity to train photonic neural networks using the nonlinear electro-optic response of a MDM. The computation speed (\textit{i.e.}, response time) is mainly limited by the optical modulation and detection bandwidths that can reach several gigahertz in commercially available silicon photonic processes. Note that the required electronics for OEO conversion can be co/hybrid integrated with the photonic chip. Our method can be scaled to a large number of neurons and layers and is an important step towards full integration of backpropagation training and inference on the same photonic chip that enables faster and more efficient photonic computing systems.

\backmatter

\bibliography{sn-bibliography}

\end{document}